\documentclass[preprint]{elsarticle}
\usepackage{amssymb}
\usepackage[T1]{fontenc}
\usepackage{mathtools}
\usepackage{breqn}
\usepackage{xfrac}
\usepackage{braket}
\usepackage{physics}
\usepackage{tikz}
\usepackage{xparse}
\usepackage{float}
\usepackage{varwidth}
\usepackage[inline]{enumitem}
\usepackage{xcolor}
\usepackage{amsthm}
\usepackage[noabbrev,capitalize]{cleveref}
\usepackage{changepage}
\usepackage[normalem]{ulem}
\usepackage{booktabs}
\usepackage{array}
\usepackage[shortcuts]{extdash}

\newcolumntype{L}[1]{>{\raggedright\let\newline\\\arraybackslash\hspace{0pt}}m{#1}}
\newcolumntype{C}[1]{>{\centering\let\newline\\\arraybackslash\hspace{0pt}}m{#1}}
\newcolumntype{R}[1]{>{\raggedleft\let\newline\\\arraybackslash\hspace{0pt}}m{#1}}

\usetikzlibrary{fit, shapes.geometric, fadings, patterns, positioning, quotes}

\tikzfading[name=fade out, inner color=transparent!0, outer color=transparent!100]{}

\newtheorem{theorem}{Theorem}
\newtheorem{lemma}[theorem]{Lemma}

\NewDocumentEnvironment{delineate}{m}{\textcolor{cyan!70!black!}{> > > > Begin: #1 > > > >}}{\textcolor{red!70!black!}{< < < < End: #1 < < < <}}

\mathchardef\mhyphen="2D

\renewcommand{\P}[1]{\mathcal{P}\paren*{#1}}

\DeclarePairedDelimiter{\ceil}{\lceil}{\rceil}

\DeclarePairedDelimiter\paren\lparen\rparen
\DeclarePairedDelimiter\aparen\langle\rangle

\newcommand\langclassformat[1]{\begingroup\ensuremath{\rm #1}\endgroup}
\newcommand\restrictionformat[1]{\begingroup\ensuremath{\tt #1}\endgroup}
\newcommand\zeroX{\restrictionformat{0}}
\newcommand\consX{\restrictionformat{con}}
\newcommand\logX{\restrictionformat{log}}

\newcommand\polyX{\restrictionformat{poly}}

\newcommand\rtX{\restrictionformat{rt}}
\newcommand\owayX{\restrictionformat{1way}}

\newcommand\Xspace[1]{\restrictionformat{#1\mhyphen\allowbreak{}space}}
\newcommand\Xrandom[1]{\restrictionformat{#1\mhyphen\allowbreak{}random\mhyphen\allowbreak{}bits}}
\newcommand\Xtime[1]{\restrictionformat{#1\mhyphen\allowbreak{}time}}
\newcommand\Xinput[1]{\restrictionformat{#1\mhyphen\allowbreak{}input}}

\newcommand{\vgeneric}[1]{\langclassformat{VER(\scalebox{0.8}{$#1$})}}
\newcommand{\vsri}[3]{\vgeneric{\Xspace{#1},\allowbreak\Xrandom{#2},\allowbreak\Xinput{#3}}}

\newcommand{\vtsr}[3]{\vgeneric{\Xtime{#1},\allowbreak\Xspace{#2},\allowbreak\Xrandom{#3}}}
\newcommand{\vsr}[2]{\vgeneric{\Xspace{#1},\allowbreak\Xrandom{#2}}}
\newcommand{\langclass}{\vsri{\consX}{\consX}{\rtX}}

\newcommand{\vconcise}[1]{\langclassformat{\mathcal{V}(\scalebox{0.9}{$#1$})}}
\newcommand{\vcscri}[1]{\vconcise{\Xinput{#1}}}

\newcommand{\ODFAK}[1]{\langclassformat{1DFA\paren*{#1}}}

\newcommand{\ONFAK}[1]{\langclassformat{1NFA\paren*{#1}}}

\newcommand{\PP}{\langclassformat{P}}
\newcommand{\NP}{\langclassformat{NP}}
\newcommand{\NL}{\langclassformat{NL}}

\newcommand{\dfa}{\ensuremath{\mathrm{dfa}}}
\newcommand{\nfa}{\ensuremath{\mathrm{nfa}}}
\newcommand{\pfa}{\ensuremath{\mathrm{pfa}}}

\newcommand{\odfa}{\ensuremath{\mathrm{1\dfa}}}
\newcommand{\odfak}[1]{\ensuremath{\odfa\paren*{#1}}}

\newcommand{\onfa}{\ensuremath{\mathrm{1\nfa}}}
\newcommand{\onfak}[1]{\ensuremath{\onfa\paren*{#1}}}

\newcommand{\onfaklnrr}[2]{\ensuremath{\ensuremath{\mathrm{1\mathrm{ns}\nfa}}\paren*{#1, #2}}}
\newcommand{\ONFAKLNRR}[2]{\ensuremath{\langclassformat{\mathrm{1\mathrm{NS}\mathrm{NFA}}}\paren*{#1, #2}}}

\newcommand{\machine}{\ensuremath{\mathrm{\pfa\mhyphen v}(\Xspace{\consX},\allowbreak\Xinput{\rtX})}}

\newcommand{\ith}[2][th]{\ensuremath{#2}\text{#1}}

\newcommand{\inpAlp}{\Sigma}
\newcommand{\certAlp}{\Gamma}

\newcommand{\certAlps}{\certAlp^{*}}

\newcommand{\langformat}[1]{\ensuremath{#1}}
\newcommand{\twin}{\langformat{L_{\mathrm{\langformat{twin}}}}}
\newcommand{\twinlong}{\ensuremath{\twin = \Set{w\#w | w \in \Set{0, 1}^*}}}
\newcommand{\nonpal}{\langformat{L_{\mathrm{\langformat{nonpal}}}}}
\newcommand{\lmatch}{\langformat{L_{\mathrm{\langformat{match}}}}}
\newcommand{\lfrac}{\ensuremath{L_{\frac23}}}

\newcommand{\qacc}{\ensuremath{q_{\mathrm{acc}}}}
\newcommand{\qrej}{\ensuremath{q_{\mathrm{rej}}}}

\newcommand{\lend}{\ensuremath{\rhd}}
\newcommand{\rend}{\ensuremath{\lhd}}

\newcommand{\Accept}{\textit{Accept}}

\newcommand{\reject}{\textit{reject}}

\newenvironment{turing}[2]
 {\begin{enumerate}[labelsep=0pt,align=left,parsep=0pt,leftmargin=0pt]
  \item[$#1={}$]``\ignorespaces#2
  \begin{enumerate}[
    nosep,
    labelsep=.5em,
    leftmargin=\widthof{$#1={}$}+1.5em,
    labelwidth=1.5em,
    label=\bfseries\arabic{*}.,
    ref=\arabic{*}
  ]}
 {\unskip''\end{enumerate}\end{enumerate}}

\newcommand{\bitem}[1]{\item\begin{adjustwidth}{1em}{0pt}\ignorespaces#1\end{adjustwidth}}
\newcommand{\bbitem}[1]{\item\begin{adjustwidth}{2em}{0pt}\ignorespaces#1\end{adjustwidth}}

\makeatletter
\def\squiggly{\bgroup \markoverwith{\lower3.9\p@\hbox{\sixly \scalebox{1.2}[0.65]{\char58}}}\ULon}
\makeatother
\newcommand{\stkout}[1]{\ifmmode\text{\sout{\ensuremath{#1}}}\else\sout{#1}\fi}
\newcommand{\utkanadd}[1]{\textcolor{blue!20!cyan!40!black}{\ifmmode\smash[b]{\squiggly{#1}}\else\squiggly{#1}\fi}}

\begin{document}

\begin{frontmatter}

\title{Real-Time, Constant-Space, Constant-Randomness Verifiers\tnoteref{confref}}
\tnotetext[confref]{This paper is a substantially improved version of~\cite{DEGS22}.}

\author{M. Utkan Gezer\corref{cor1}}
\ead{utkan.gezer@boun.edu.tr}

\author{\"{O}zdeniz Dolu}
\ead{ozdeniz.dolu@boun.edu.tr}

\author{Nevzat Ersoy}
\ead{nevzat.ersoy@boun.edu.tr}

\author{A. C. Cem Say}
\ead{say@boun.edu.tr}

\cortext[cor1]{Corresponding author}
\address{Department of Computer Engineering, Bo\u{g}azi\c{c}i University, Bebek 34342, \.{I}stanbul, Turkey}

\begin{abstract}
    We study the class of languages that have membership proofs which can be verified by real-time finite-state machines using only a constant number of random bits, regardless of the size of their inputs. Since any further restriction on the verifiers would preclude the verification of nonregular languages, this is the tightest  computational budget which allows the checking of  externally provided proofs  to have meaningful use.  
    We provide a full characterization of this class of languages in terms of a restricted version of the one-way nondeterministic multihead finite automaton model.
    For any $k>0$, there exist languages that cannot be recognized by any $k$\=/head one-way nondeterministic  finite automaton, but that are nonetheless real-time verifiable in this sense.
    The set of nonpalindromes, which cannot be recognized by any one-way multihead deterministic finite automaton, is also demonstrated to be verifiable within these restrictions. 
    
\end{abstract}


\end{frontmatter}

\section{Introduction}
The characterization of problem classes in terms of the computational requirements on machines that are supposed to check purported proofs of membership of their input strings in a language has been an important theme of complexity theory, leading to landmark achievements like the PCP Theorem~\cite{ALMSS98,AS98} and celebrated open questions like the \PP{} vs.\ \NP{} problem. 

As expected, imposing tighter bounds on the computational resources of the verifiers for these proofs of membership seems to restrict the associated language classes: Limiting a polynomial-time deterministic verifier to use only a logarithmic, rather than polynomially bounded amount of working memory ``shrinks'' the class of verifiable languages to \NL{} from \NP{}, and the same apparent loss of power also occurs when a logarithmic-space, polynomial-time probabilistic verifier is restricted to use only a constant, rather than logarithmically bounded number of random bits for bounded-error verification.~\cite{CL95,SY14} 

In this paper, we focus on the tightest possible ``budgetary'' restrictions that can be imposed on such verifiers by considering the  case where the machine's working memory \textit{and} the amount of usable random bits are both constants irrespective of the input length, and the runtime is maximally constrained, so that only a real-time scan of the input string is allowed. We examine the class of languages whose membership proofs can be checked under these extreme conditions. Note that decreasing the number of random bits from a positive constant to zero would make such a  proof system equivalent to a nondeterministic finite automaton, unable to recognize any nonregular languages. Since membership in any regular language can be decided by  a ``stand-alone'' real-time deterministic finite automaton with no need of an externally provided  certificate, the machines we consider are truly  the weakest possible verifiers of meaningful use, highlighting the issues involved in the checking of the proofs of extremely long claims.

We build on previous work~\cite{SY14} which showed an equivalence between con\-stant-space, constant-randomness verifiers and multihead nondeterministic finite automata working as language recognizers. This equivalence breaks down when the machines are restricted to consume their inputs in real-time fashion: A
multihead automaton
that is forced to move all of its heads forward at every step
is no more powerful than a single-head one, and can only recognize a regular language, whereas Say and Yakary\i{}lmaz were able to demonstrate a nonregular language~\cite{SY14} which has membership proofs that can be checked by a real-time finite-state verifier with a fixed number of coin tosses. 

In this paper, we 
give a full characterization of the class of languages that have membership proofs that can be verified by these very weak machines.
It turns out that this class corresponds precisely to those languages that can be recognized by a restricted model of one-way nondeterministic multihead finite automata which are forced to move their heads in turns according to a predefined order, and which make at most one nondeterministic choice for each such round of moves.
All  two-head one-way deterministic finite automata 
can be made to satisfy these conditions without changing the recognized language.
For any $k>0$, there exist languages that cannot be recognized by any $k$\=/head one-way nondeterministic  finite automaton, but that are nonetheless real-time verifiable in this sense.
The set of nonpalindromes, which cannot be recognized by any one-way multihead deterministic finite automaton, is also demonstrated to be  verifiable in this setup. We conjecture that the real-time requirement truly decreases the verification power, i.e., that there exist languages that can be verified only when the definition of these machines is relaxed to allow them the ability  to pause on the input tape.  

The rest of the paper is structured as follows: \Cref{sec:prel} provides the necessary definitions and previous results regarding the relation between multihead finite automata and constant-randomness finite-state verifiers. Our results are presented in \cref{sec:results}. \Cref{sec:conc} is a conclusion.





\section{Preliminaries}\label{sec:prel}

\subsection{One-way multihead finite automata}\label{subs:nfaks}
A \textit{one-way $k$\=/head nondeterministic finite automaton} (\onfak{k}) is a nondeterministic finite-state machine with $k$ read-only heads that it can direct on an input string flanked by two end-marker symbols. Each head can be made to stay put or move one symbol to the right in each computational step.  Formally, a \onfak{k}  is a 6-tuple $(Q,\Sigma,\delta,q_0,\qacc,\qrej)$, where
\begin{enumerate}
    \item $Q$ is the finite set of internal states,
    \item $\Sigma$ is the finite  input alphabet,
    \item $\delta \colon Q \times \Sigma_\bowtie^k \to \P{Q \times \Delta^k}$ is the transition function describing the sets of alternative moves the machine may perform at each execution step, where each move is associated with a  state to enter and whether or not to move each head, given the machine's current state and the list of symbols that are currently being scanned by the $k$ input heads:
    \begin{itemize}
        \item $\Delta = \Set{0, +1}$ is the set of possible head movements, where  $0$ means ``stay put'' and $+1$ means ``move right'',
        \item $\Sigma_\bowtie = \Sigma \cup \Set{\lend, \rend}$, where $\lend, \rend \notin \Sigma$ are respectively the left and  right end-markers, placed automatically to mark the boundaries of the input,
    \end{itemize}
    \item $q_0 \in Q$ is the  initial state,
    \item $\qacc \in Q$ is the final state at which the machine halts and accepts, and
    \item $\qrej \in Q$ is the final state at which the machine halts and rejects.
\end{enumerate}

Given an input string $w \in \Sigma^*$, a \onfak{k} $M = \paren*{Q, \Sigma, \delta, q_0, \qacc, \qrej}$  begins execution from the state $q_0$, with $\lend w \rend$ written on its tape, and all $k$ of its heads on the left end-marker.  At each timestep, $M$ nondeterministically updates its state and head positions according to the choices dictated by its transition function. Computation halts if one of the states \qacc{} or \qrej{} has been reached, or a head has moved beyond the right end-marker.

Each different sequence of choices $M$ may take corresponds to a different \textit{computation history}, i.e., a  sequence of tuples describing all the state and head positions that $M$ goes through in that particular eventuality.

$M$ is said to \emph{accept} $w$ if there exists a computation history where it reaches  the state \qacc{}, given $w$ as the input.  $M$ is said to \emph{reject} $w$ if every computation history of $M$ on $w$ either reaches  \qrej{}, ends with a  transition whose associated set of choices is $\emptyset$, or a head has moved beyond the right end-marker without a final state being entered.  $M$ might also loop on the input $w$, neither accepting nor rejecting it.

The \textit{language recognized by $M$} is the set of strings that it accepts.
%
%

A \textit{one-way $k$\=/head deterministic finite automaton}, denoted \odfak{k}, is a  \onfak{k} $(Q, \Sigma, \delta, q_0, \qacc, \qrej)$ whose transition function presents exactly one ``choice'' of move for every input ($\abs{\delta\paren*{q, x_1, \dotsc, x_k}} = 1$ for all $q \in Q$ and $x_1, \dotsc, x_k \in \Sigma_\bowtie$).  


\onfak{1} and \odfak{1} are simply called \textit{one-way nondeterministic} and \textit{deterministic finite automata}%
, respectively.  
The ``real-time'' versions of  these single-head machines are obtained by forcing the head to move to the right at each step (by setting $\Delta = \Set{+1}$). Real-time nondeterministic and deterministic finite automata
have runtimes of at most $n+2$ on input strings of length $n$.


The classes of languages recognized by each of the machine models defined above will be denoted by the uppercase versions of the associated machine denotations. For example, \ONFAK{6} denotes the class of languages recognizable by \onfak{6}'s.
The following facts~\cite{HKM11} about these language classes will be useful:

For any $k\geq 1$,
\begin{align*}
    \ODFAK{k} &\subsetneq \ODFAK{k+1}.\\
    \ONFAK{k} &\subsetneq \ONFAK{k+1}.
\end{align*}

\nonpal{} is the language which contains every string except palindromes on the alphabet $\Set{0,1}$. This language can be recognized by a \onfak{2}. Since its complement  cannot be recognized by any \odfak{k} for any $k$~\cite{KMW16}, there exists no deterministic one-way multihead automaton that recognizes \nonpal{} either, by the fact~\cite{R66} that the class of languages recognized by \odfak{k}'s is closed under complementation. This proves the inequality
$\ONFAK{2} \setminus \bigcup_k \ODFAK{k} \neq \emptyset$.

\subsection{Verifiers}\label{subs:vers}

There exist several elegant characterizations of language classes in terms of bounds imposed on the resources available to probabilistic Turing machines (``verifiers'') tasked with checking purported proofs (``certificates'') of membership of their input strings in a language. 

Formally, a 
\textit{verifier}
is a 6-tuple $(Q,\Sigma,\Phi,\Gamma,\delta,q_0)$, where
\begin{enumerate}
    \item $Q$ is the finite  set of states, such that $Q = P \cup D \cup \Set{\qacc, \qrej}$ where 
    \begin{itemize}
        \item $P$ is the set of coin-tossing states,
        \item $D$ is the set of deterministic states, such that $P \cap D = \emptyset$, and
        \item \qacc{} and \qrej{} are the accept and reject states,  respectively.
    \end{itemize}
    \item $\inpAlp$ is the input alphabet, not containing the end-markers \lend{} and \rend{},
    \item $\Phi$ is the work tape alphabet,
    \item $\certAlp$ is the  certificate alphabet, not containing 
    \rend,
    \item $\delta$ 
    is the transition function, described below, and
    \item  $q_0$ is the initial state, $q_0 \in Q$.
\end{enumerate}

As in \cref{subs:nfaks}, $\Sigma_\bowtie$ will be used to denote the union $\Sigma \cup \Set{\lend, \rend}$.

The transition function $\delta$ is constructed in two parts, as follows: For $q \in P$, $q' \in Q$, $\sigma \in \Sigma_\bowtie$, $\phi, \phi' \in \Phi$, $\gamma \in \Gamma\cup \Set{
\rend}$, $b \in \Set{0,1}$, $d_i, d_w \in \Set{-1,0,+1}$, and $d_c \in \Set{0,+1}$,  $\delta(q,\sigma,\phi,\gamma,b)=(q',\phi',d_i,d_w,d_c)$ dictates that the machine will switch to state $q'$, write $\phi'$ on the work tape, and move the input, work tape and certificate heads in directions $d_i$, $d_w$, and $d_c$, respectively, if it is originally in state $q$, scanning $\sigma$, $\phi$, and $\gamma$ on the three respective tapes, and has obtained the random bit $b$ as the result of a fair coin toss. For $q\in D$, $\delta(q,\sigma,\phi,\gamma)=(q',\phi',d_i,d_w,d_c)$ dictates a similar, but deterministic transition.

A verifier halts with acceptance (rejection) when it executes a transition entering \qacc{} (\qrej). Any transition that  moves the input or certificate head  beyond an  end-marker delimiting the  string written on the associated  read-only tape leads to a rejection, unless that last move enters \qacc. 
The head on the  certificate tape is defined to be one-way, since it is known~\cite{AB09} that allowing two-way access to that tape can lead to ``unfair'' accounting of the space usage. The input and work tape heads are two-way in the general definition above, although we will be considering restricting the  movement types of  the input tape head (and completely removing the work tape) in most of the following. 

We say that such a machine $V$ verifies a language $L$ with error $\epsilon$ if there exists a number $\epsilon<1$ where
\begin{itemize}
    \item for all input strings $w \in L$, there exists a certificate string $c_w$ such that $V$ halts by accepting with probability 1 when started on $w$ and $c_w$, and,
    \item for all input strings $w \notin L$ and for all certificates $c$, $V$ halts by rejecting with probability at least $1-\epsilon$ when started on $w$ and $c$.
\end{itemize}

We will be using the notation \vgeneric{\restrictionformat{restriction}_1, \restrictionformat{restriction}_2, \dotsc, \restrictionformat{restriction}_k} to denote the class of languages that can be verified by machines that, on all of their possible computations, halting or non-halting, operate within the added restrictions indicated in the parentheses. These may represent  bounds for runtime, working memory usage, and number  of random bits to be used as a function of the length of their input strings. The terms \zeroX, \consX, \logX, and \polyX{} will be used to represent the well-known types of functions to be considered as resource bounds, with ``\consX'' standing for constant functions of the input length, the others being self evident, to form arguments like ``\Xtime{\polyX}'' or ``\Xspace{\logX}''. The ``one-way'' mode, where the input  head is not allowed to move left, will be indicated by the parameter ``\Xinput{\owayX}'', whereas the further restriction to real-time movement, where the head is not allowed to pause at any step during its left-to-right scan, will be indicated by ``\Xinput{\rtX}''. 

The following characterizations in terms of zero-error verifiers are well known:
\begin{align*}
    \vtsr{\polyX}{\polyX}{\zeroX} &= \NP\\
    \vtsr{\polyX}{\logX}{\zeroX} &= \NL
\end{align*}

When one allows nonzero error, significant gains in space usage seem to be achievable:
\begin{align*}
    \vtsr{\polyX}{\logX}{\logX} &= \NP \tag*{\cite{CL95}}\\
    \vsr{\consX}{\consX} &= \NL \tag*{\cite{SY14}}
\end{align*}

For verifiers using at least logarithmic space, the magnitude of the one-sided error can be reduced without significant increase in the runtime, whereas the constant-space verifiers of~\cite{SY14} (all of which have correct certificates that can be checked  in polynomial time) do not seem~\cite{GS21} to have this property in general.\footnote{Note that a constant-space machine  is equivalent to a finite-state automaton with no work tape, since the bounded amount of information in the work tape of a constant-space verifier can also be kept using a suitably large set of internal states.}

Say and Yakary\i{}lmaz~\cite{SY14} also considered the case where a constant-space, constant-randomness verifier is forbidden to move its input head to the left. 
In the rest of this paper, we will be focusing on this model and its further specialization where the input head is forced to move at each step. We will use the following less cumbersome notation for the associated language classes:
\begin{align*}
    \vcscri{\owayX} &:= \vsri{\consX}{\consX}{\owayX}\\
    \vcscri{\rtX} &:= \vsri{\consX}{\consX}{\rtX}
\end{align*}

Using the techniques introduced in~\cite{SY14}, one can obtain the following characterization:

\begin{theorem}\label{thm:SY}
$\vcscri{\owayX} = \bigcup_k \ONFAK{k}$.
\end{theorem}
\begin{proof}

Given a \onfak{k} $M$ recognizing a language $L_M$, one can construct a one-way, constant-space, constant-randomness verifier $V_M$ for $L_M$ as follows: $V_M$ expects the certificate to contain a proof of the existence of an accepting computation history (in the form of a sequence of tuples representing the nondeterministic branch taken and list of symbols scanned by the heads at each step) of $M$ working on the input string. $V_M$ uses its random bits to select a head of $M$ and simulates its execution on the input, relying on the certificate for information on what symbols would be scanned by the other heads of $M$ at every step. If $V_M$ ever sees the certificate reporting that the head it is tracking is currently scanning a symbol other than the correct value, it rejects. If the input is in $L_M$, a correct certificate that carries $V_M$ to acceptance with probability 1 exists. Otherwise, in order to trick $V_M$ to reach an accept state, the certificate would have to ``lie'' about what is being seen by at least one of the heads of $M$ in at least one step, and $V_M$ has a constant probability of having selected that head, and therefore rejecting the input. Since $M$ can be assumed to run in linear time in all its nondeterministic branches without loss of generality, any attempt by an overly long certificate to trick $V_M$ to loop without accepting will also be caught by nonzero probability.

In the reverse direction, given a finite-state verifier $V$ with one-way input that uses at most $r$ random bits,
one can build a \onfak{2^r} $M_V$ for the verified language $L_V$ as follows: $V$'s behavior on each different random bit sequence can be represented by a deterministic verifier obtained  by ``hardwiring'' that particular sequence into $V$'s transition function. $M_V$ is designed to nondeterministically guess a certificate and use its heads to simulate all these $2^r$ deterministic verifiers operating on the input string and the common certificate. For each newly guessed certificate symbol, $M_V$ goes through all the deterministic verifiers one by one, tracing each one's execution (by changing its state and possibly moving the corresponding head) until that deterministic verifier accepts, rejects, is about to step beyond the right end-marker on the input tape\footnote{Whenever $M_V$ ``realizes'' that the deterministic verifier it is simulating is about to take such a step, it keeps the corresponding head in place and switches to the simulation of the next deterministic verifier in that very step.
The only exception to this is the case where all but one of the deterministic verifier simulations have halted, in which case $M_V$ allows the corresponding head to  ``spill over''.}
(in which case it will be interpreted as having made the same decision that it would have made at that final step), or performs a transition consuming that new certificate symbol by moving its certificate tape head.\footnote{
$M_V$ can detect when a deterministic verifier enters an infinite loop that does not move  any  heads, and reject on such nondeterministic branches.
}
This procedure continues until either a deterministic verifier rejects, or all the $2^r$ deterministic verifiers are
seen to accept.
$M_V$ accepts if it arrives in a state representing all the deterministic verifiers having accepted.
\end{proof}

This link between finite-state constant-randomness verifiers and multihead automata is broken when one further restricts the input heads to be real-time: A multihead finite automaton operating all its heads in real time is easily seen to be no stronger than a single-head finite automaton, and therefore cannot recognize a nonregular language. Say and Yakary\i{}lmaz, however, were able to demonstrate~\cite{SY14} a finite-state constant-randomness verifier with real-time input that verifies the nonregular language \twinlong{} {on the alphabet $\Set{0,1,\#}$}:\footnote{Note that the construction in the proof of \cref{thm:SY} produces a multihead automaton with  heads that can pause on the input, even when it is fed a verifier with real-time input.} The certificate is expected to consist of the string $w$, which is supposed to appear on both sides of the symbol $\#$ in the input. The machine tosses a coin to decide whether it should compare the substring appearing to the left or to the right of the  $\#$ with the certificate as it is consuming the input in real time, and accepts only if this comparison is successful. Acceptance with probability 1 is only possible for members of the language associated with well-formed certificates.

Note that such a machine must use its capability to pause the \textit{certificate} tape head for some steps. This is easy to see when one considers the computational power of a verifier with real-time heads on both the input and certificate tapes: All the ``deterministic'' verifiers that can be obtained from the probabilistic verifier by hardwiring the possible random sequences (as we saw in the proof of \cref{thm:SY}) would then be running both their heads on exactly the same strings in perfect synchrony, and it would be possible to build a single real-time one-head finite automaton simulating this collection. This machine would be equivalent to a one-head nondeterministic finite automaton, with no power of recognizing nonregular languages.  

In a very real sense, \vcscri{\rtX} corresponds to the weakest computational setup where  externally provided proofs  are meaningful.  In the next section, we will examine this interesting class and its relationships with \vcscri{\owayX} and many of its subsets in detail.


\section{Real-time, finite-state, constant-randomness verification}\label{sec:results}

In this section, we will present a characterization of the class of languages that have proofs of membership that are verifiable by real-time, finite-state, constant-randomness machines in terms of a naturally restricted version of  the one-way nondeterministic multihead automaton model.

We note that the proof of \Cref{thm:SY} provides a method to convert any one-way nondeterministic multihead automaton to an equivalent machine in a ``canonical form'': Given some machine $M$, one can first convert it to a verifier $V$ for the same language using the first construction described in that proof. When $V$ is converted back to an equivalent \onfak{k} $M'$ using the second construction, $M'$ is guaranteed to have the following properties:

\begin{itemize}
    \item The heads ``move'' alternately according to a predefined order: The machine spends one or more steps for the first head (during which it may or may not move it forward for one or more cells on the input tape), then switches to the second head, then the third, and so on, switching back to the first head after the \ith{k} one. 
    Heads that reach the end of the input tape\footnote{When $k-1$ heads have already reached the end of the input tape, the last one is allowed to move beyond the right end-marker in its final step.} may be considered to be ``dropped out'' of this cycle,\footnote{\label{fn:consumebeforeaccept}Strictly speaking, the machine $M_V$ described in the proof of \Cref{thm:SY} can stop a head corresponding to a deterministic verifier that has accepted permanently without having to reach the end of the input tape, but every such machine can be transformed into one that moves every such head to the end of the tape.} in which case this process goes on with the remaining heads, and
    \item the automaton makes at most 
    one nondeterministic choice during each iteration of this loop.
\end{itemize}

As explained above, this is what one obtains if one converts an arbitrary constant-randomness finite-state verifier with one-way input to a \onfak{k} using the technique of \Cref{thm:SY}. When one submits a \textit{real-time} verifier to that procedure, though, the output machine satisfies an additional restriction: It moves every head (that has still not dropped out of the cycle)  forward for at least one step on the tape during each iteration of the loop. This is because each ``deterministic verifier'' used in the procedure described in that proof is a real-time automaton in this case, and simulating every transition of such a machine involves moving the input head. We will show that ``canonical form''  \onfak{k}'s with this additional property 
correspond exactly to real-time, constant-space, constant-randomness verifiers.

Let us define a slightly more powerful-looking model that will turn out to be equivalent to the restricted machines described above. For any positive integer $m$, a \textit{one-way $k$\=/head nonstop 
finite automaton} \emph{with (up to) $m$ nondeterministic choices per round} (\onfaklnrr{k}{m}) is a \onfak{k} such that
\begin{itemize}
    \item exactly one head moves at every transition,
    \item the heads which have not yet reached the end of the input tape 
    take turns in moving for one or more steps in a round-robin fashion (with the very last head to move also allowed to spill beyond the right end-marker in its last step), and
    \item the automaton makes at most $m$ nondeterministic choices during each iteration (``round'') of the head movements loop.
\end{itemize}

In keeping with our naming conventions, the class of languages recognized by some \onfaklnrr{k}{m} for any value of $k$ will be denoted  \ONFAKLNRR{*}{m}. When $m$ is also relaxed we name the corresponding class  \ONFAKLNRR{*}{*}.

\subsection{$\vcscri{\rtX} = \ONFAKLNRR{*}{1} = \ONFAKLNRR{*}{*}$}

We will show that every language recognizable by a \onfaklnrr{k}{m} is verifiable by a constant-space, constant-randomness verifier that scans its input in real time, and vice versa. One direction of this equivalence has, in fact, already been noted above. We state it below without proof for the sake of completeness:

\begin{lemma}\label{lem:vertonfak}
$\vcscri{\rtX} \subseteq \ONFAKLNRR{*}{1}$.
\end{lemma}

It remains to show that every language recognizable by a \onfaklnrr{k}{m} has a real-time verifier of the kind considered in this paper.
The technique employed in the proof of \cref{thm:SY} for constructing verifiers is not useful here, since it requires the verifier to pause its input head occasionally when processing certain portions of the certificate. We will show that, for all $m>0$, all languages in \ONFAKLNRR{*}{m} have more concise proofs of membership that can be checked by our restricted machines. 

\begin{lemma}\label{lem:nfaktover}
$\ONFAKLNRR{*}{*} \subseteq \vcscri{\rtX}$.
\end{lemma}

\begin{proof}
\newcommand{\sh}{\ensuremath{H}}
\newcommand{\shn}{\ensuremath{m}}
\newcommand{\finald}{\ensuremath{d'}}
\newcommand{\finali}{\ensuremath{i'}}
Let $M = \paren{Q_M, \Sigma, \delta_M, q_0, \qacc, \qrej}$ be a \onfaklnrr{k}{m} recognizing the language $A$.  Knowing about the way that $M$ moves its heads, a computation history of $M$ can be viewed as the concatenation of infinitely many  sub-histories
$\sh_{1,1},\allowbreak \sh_{1,2}, \dotsc,\allowbreak \sh_{1,k},\allowbreak \sh_{2,1},\allowbreak \sh_{2,2}, \dotsc$, where only the \ith{i} head moves during 
$\sh_{d,i}$ (where $d>0$ is the round index), and $\sh_{d,i}$ is empty either if the \ith{i} head has already reached the end of the input or if $M$ has already halted by that time. Let $\sh_{\finald, \finali}$ be the halting (accepting or rejecting) sub-history at the end of which $M$ accepts or rejects, and beyond which all the sub-histories are empty.  Let us call 
$\sh_{1,i}\sh_{2,i}\sh_{3,i}\dotsm$, the concatenation of all the sub-histories corresponding to the \ith{i} head, the \emph{\ith{i} part} of the history.

Note that, if one visualizes the \ith{i} part of the history for any specific $i$, one sees the \ith{i} head moving in real time on the input until $M$ halts. Furthermore, the state sequence traversed during these moves is easy to trace
for a real-time verifier that employs
$M$'s transition function and that has been given a list of the nondeterministic choices made by $M$, except at the ``joints'' between sub-histories, where the machine's state and the positions of the other heads make ``leaps'' corresponding to (possibly long) sequences of moves carried out by those other heads while the \ith{i} head was pausing.
Intuitively, each part of the history can be thought of as describing the execution of a real-time automaton that momentarily ``blacks out'' as it switches from any 
$\sh_{d,i}$ to $\sh_{(d+1),i}$, finding the machine's state and the other heads' positions updated to new values when it ``wakes up.''  Our strategy for real-time verification will follow directly from this observation, and the certificate will supply the necessary information to deal with the blackouts.

\newcommand{\lst}{\ensuremath{s}}
\newcommand{\lsy}{\ensuremath{z}}
\newcommand{\lsn}{\ensuremath{y}}

We will construct a real-time, finite-state verifier $V$ that uses $r = \ceil{\log_2(k)}$ random bits to verify proofs of membership in the language $A$. $V$ will use these random bits to choose a  head of $M$, say, the \ith{i} one,  in private, 
and trace the \ith{i} part of $M$'s (purportedly accepting) computation history.

The certificate $c_w$ for a string $w \in A$ is expected to present the succinct sequence of updates needed for $V$ to recover from the aforementioned blackouts.
To serve that purpose, and to also provide $V$ with the nondeterministic choices of $M$, the certificate alphabet of $V$ will be $\paren*{Q_{M} \times \Sigma_\bowtie  \times Y^m}^k$, 
where $Y$ is the set of nondeterministic choices (e.g., $Q_M \times \Delta^k$),
and
the \ith{d} symbol $c_{w,d}$  of the certificate will look like 
%
%
%
\begin{gather*}
    c_{w,d} = \paren*{c_{w,d,1}, c_{w,d,2}, \dotsc, c_{w,d,k}},\\
    \intertext{where each sub-symbol $c_{w,d,i}$ is}
    c_{w,d,i} = \paren[\Big]{\lst_{d,i}, \lsy_{d,i}, \paren*{\lsn_{d,i,1}, \lsn_{d,i,2}, \dotsc, \lsn_{d,i,m}}}.
\end{gather*}
The certificate's ``claims'' about $M$'s computation that have been encoded into symbols like $c_{w,d}$ are as follows: For each $d, i > 0$, $M$ will be in state 
$\lst_{d, i}$, and the \ith{i} head will be scanning the input symbol
$\lsy_{d, i}$ at the end of
$\sh_{d, i}$.
For any specific pair $d, i > 0$, some prefix of $\paren*{\lsn_{d,i,1}, \dotsc, \lsn_{d,i,m}}$ indicates the nondeterministic choices that $M$ would make in its \ith{d} round during the \ith{i} head's turn to stay on a path leading to  acceptance.
The rest of the $\lsn_{d, i,j}$'s are to be ignored.


$c_w$ concludes its claim that $M$'s computation accepts at $\sh_{\finald,\finali}$ (with $\lst_{\finald,\finali} = \qacc$) by the final symbol $c_{w,(\finald+1)}$. The sole function of the sub-symbols beyond $c_{w,\finald,\finali}$ is to inform the branches of $V$ that follow the \ith{j} head (for $j \neq i$) that $M$ has already accepted, as will be described below.


For the sake of simplifying the exposition ahead, let
\begin{equation*}
\begin{aligned}
    \lst_{1, 0} &= q_0,\\
    \lst_{d, 0} &= \lst_{(d-1), k} &&\text{for $d > 1$, and}\\
    \lsy_{0, i} &= \lend           &&\text{for $0 < i \leq k$}.
\end{aligned}
\end{equation*}

Given an input $w$ and a certificate $c_w$, $V$ starts by tossing  $r$ coins to choose which head of $M$ to trace. (By definition, $V$ will have to move its input head while tossing these coins.  We will first explain the general algorithm as if $V$ can hold its input head in place until this coin-tossing stage is over, and later elaborate on how this behavior can be simulated by a real-time verifier.) 
$V$ will scan the input in real time in an attempt to verify the claims of the  certificate about the  part of $M$'s computation history corresponding to the chosen (say, \ith{i}) head of $M$ by simulating $M$'s actions in that part. 

For each $d > 0$, $V$ starts its simulation of the sub-history $H_{d,i}$ (from the state
$\lst_{d,(i-1)}$) with its certificate head scanning the symbol
$c_{w,d}$. If the certificate indicates that $M$ has already halted 
during some sub-history prior to $\sh_{d, i}$, $V$ should halt with the same verdict at this step. 
Therefore, $V$ accepts (rejects) immediately if $\lst_{d, (i-1)}$ equals \qacc{} 
(\qrej{}).
If the certificate claims that $M$ has not halted yet, $V$ 
memorizes $c_{w,d}$ in its internal state and 
moves the certificate head forward in this first step (simultaneously with the always-moving input head) to scan $c_{w,(d+1)}$, which contains information that will  be needed at the instant when this sub-history simulation comes to an end.     
During this simulation, $V$ assumes that
every other head (``the \ith{j} head'' for any $j \neq i$)  is paused on the symbol
\begin{equation*}
\begin{aligned}
     &\lsy_{d, j}     &&\text{for } j<i,\\
     &\lsy_{(d-1), j} &&\text{for } j>i,
\end{aligned}
\end{equation*}
and traces the \ith{i} head on its own in real time. If  it reaches one or more nondeterministic branching points for $M$ during the simulation of  
the sub-history $\sh_{d, i}$, $V$
takes the paths indicated by $\lsn_{d,i,1}, \dotsc, \lsn_{d,i,m}$ in the order they are provided, rejecting if they are invalid at that point of the simulation.
If it faces an additional nondeterministic branching of $M$ for which the certificate has not provided a direction, $V$ again rejects.\footnote{
The certificate must contain a ``lie'' for $V$ to be faced with such a situation, and another probabilistic branch of the verifier is guaranteed to reject upon detecting that lie, as will be noted shortly.
}



$V$ maintains a list of the heads of $M$ that have previously been claimed by the certificate to have reached the end of the input. For each such head indexed by $j > 0$, $V$ checks whether the information in $c_{w,d}$ is consistent with that list, i.e., that  $\lsy_{d, j} = \rend$ and $\lst_{d, j} = \lst_{d, (j-1)}$, rejecting immediately if it discovers an inconsistency. It also checks whether any new head should be added to the list by checking whether $\lsy_{d, j}=\rend$ for all other $j$.

\newcommand{\lsyc}{\ensuremath{\check{\lsy}}}
\newcommand{\lstc}{\ensuremath{\check{\lst}}}


If $V$ reaches  a halting state of $M$ during its simulation, it halts with the same decision.  Otherwise, as $\sh_{d,i}$ comes to an end, the simulated machine $M$ would pause the \ith{i} head and move some other head, whereas $V$ cannot pause its real-time head.  When $V$ sees that its simulation of $M$ has reached such a juncture at state $\lstc$ while its input head is scanning the symbol $\lsyc$, it does the following two things at once in the next step:
\begin{enumerate}
    \item $V$ verifies that $\lstc = \lst_{d, i}$ and that 
    $\lsyc=\lsy_{d, i}$ (in line with the claims of 
    $c_{w,d}$)
    and rejects immediately if it discovers an inconsistency.
    \item
    If 
    $\lsyc = \rend$, $V$ accepts. Otherwise, it starts a simulation of the sub-history $H_{(d+1),i}$ from the state $\lst_{(d+1),(i-1)}$ with its certificate head scanning the symbol $c_{w,(d+1)}$, repeating the process described above.
\end{enumerate}

It is important to note that the definition of a real-time verifier necessitates $V$ to move its input head while tossing its $r$ coins, and also that we can alleviate this issue:  $V$ keeps these first $r$ input symbols in its finite-state storage.  Coinciding with its latest coin flip, $V$ processes all these memorized input symbols at once in a single macro-step that is merely a concatenation of the individual steps described above.  $V$ may also need to know about the first $r$ certificate symbols while taking that macro-step.  To help with that, as with the input symbols, $V$ passes over and memorizes the first $r$ certificate symbols while tossing its coins, and uses as many of them as it needs during the macro-step.   During the rest of its simulation, $V$ continues to consume this memorized certificate prefix until it is exhausted, and then proceeds to read  the rest of certificate.

The number of input strings shorter than the constant $r$ are finite.  Therefore, if $V$ reaches the right-end marker on the input before the coin-tossing stage is over, it 
can immediately decide upon the membership of that short input string with zero error by just using a look-up table.


The reader might have observed that 
$V$'s algorithm, as described, is able to simulate  \onfak{k}'s which have an even higher ``nondeterminism budget'' than \onfaklnrr{k}{m}'s, namely, those which make at most  $m$ nondeterministic choices during each individual head's movement turn in each round of their executions.
Such an event can never occur in any real execution of $M$ by definition.  For this to arise during $V$'s simulation, the certificate must have lied at some point prior to that event, which would be caught by the probabilistic branch that is verifying the respective part of the history, as described below.

Finally, we note that $V$ can use information written in up to two certificate symbols while simulating $M$ during a particular sub-history: The last two symbols ($c_{w,(d-1)}$ and $c_{w,d}$) to the left of the certificate head are kept in the machine's finite-state storage for this purpose.


Each probabilistic branch of $V$ accepts if and only if their simulation of $M$ reaches an accept state of $M$ or its input head reaches the right end-marker without rejecting.
$V$ accepts all $w \in A$ with probability 1 when coupled with a certificate $c_w$ describing the sub-history transitions correctly for at least $\finald+1$ rounds. Whenever $w \notin A$, a $c_w$ that describes any computation history of $M$ faithfully will lead all branches of $V$ to rejection.
Therefore, the first branch of $V$ to accept a $w \notin A$ must have been brought there by a lie regarding either a $\lsy_{d, i}$ or a $\lst_{d, i}$ before that point, which will be caught out by the branch that has direct access to the relevant state and head information (i.e., the branch that picks the \ith{i} head to trace).
As a result, all nonmembers of $A$ will be rejected with probability at least $\sfrac{1}{2^r}$.
%
%
%
%
%
%
\end{proof}


\Cref{lem:vertonfak,lem:nfaktover} constitute the proof of our main result:
\begin{theorem}\label{theorem:nfakequalsver}
$\vcscri{\rtX} = \ONFAKLNRR{*}{1} = \ONFAKLNRR{*}{*}$.
\end{theorem}

\subsection{Relationships with other subsets of $\bigcup_k \ONFAK{k}$}
\label{subsec:beyond1dfa2}

It is instructive to examine several concrete examples of languages that are real-time verifiable by constant-randomness, constant-space machines. \Cref{table:dfa2} lists several  members of the class \ODFAK{2} that are easily seen to be in \vcscri{\rtX}, together with certificate templates for members of each example language.\footnote{We use the common alphabet $\Set{0,1,\#}$ for these languages. Note that the complements of all the languages listed in \Cref{table:dfa2} are also in \vcscri{\rtX}.} 
This is not surprising, as the proof of the following theorem, provided for completeness, is folklore:

\newlength{\colpadding}

\begin{table}
\caption{Certificate templates for some members of  \ODFAK{2}.}\label{table:dfa2}
\centering
\aboverulesep=0ex
\belowrulesep=0ex

\renewcommand{\arraystretch}{1.42}
\setlength{\colpadding}{6pt}
\newcommand{\colstartendmultiplier}{0.7}

\begin{tabular}{@{\hspace{\colstartendmultiplier\colpadding}}L{13.95em}@{\hspace{\colpadding}} | @{\hspace{\colpadding}}L{11.75em}@{\hspace{\colpadding}} | @{\hspace{\colpadding}}C{5.3em}@{\hspace{\colstartendmultiplier\colpadding}}} \toprule
\textbf{Language} & \textbf{Archetypal member $w$} & \textbf{Certificate for $w$} \\ \midrule
\twin & $x\#x$, where $x \in \Set{0, 1}^*$ & $x$ \\ \midrule
The set of all strings containing equal numbers of 0's and 1's & $x$, where $x$ contains $m$ 0's and $m$ 1's & $1^m$ \\ \midrule
The set of all odd-length  binary strings with the symbol~$\#$ at the middle position & $x\#y$, where $x, y \in \Set{0, 1}^*$ and $\abs{x}=\abs{y}$ & $1^{\abs{x}}$ \\ \midrule
$\Set{w | w \in (x\#)^+, x \in (0\cup 1)^+}$ & $(x\#)^m$, where $x \in (0\cup 1)^+$ & $(x\#)^{m-1}$ \\ \bottomrule
\end{tabular}
\end{table}





\begin{theorem}\label{lem:1dfak2subseteq1nfaklnrr}
$\ODFAK{2} \subseteq \ONFAKLNRR{*}{1}$.
\end{theorem}

\begin{proof}
%
Let $M = \paren{Q_M, \Sigma, \delta_M, q_0, \qacc, \qrej}$ be a \odfak{2}.  At any step of its execution, $M$ might be moving none, one, or both of its heads.  We can modify $M$ to obtain a \odfak{2} $M' = \paren{Q_{M'}, \Sigma, \delta_{M'}, q_0, \qacc, \qrej}$ that recognizes the same language while moving exactly one of its heads at every step, starting with the first head.  With two heads moving in an alternating fashion, the resulting machine $M'$ will then also be a \onfaklnrr{2}{0} by definition (which is also a \onfaklnrr{2}{1}, again by definition).  To complete the proof, the details of $M'$'s construction is as follows:

The state set of the machine $M'$ is defined as  $Q_{M'} = Q_M \cup \Set{q' | q \in Q_M}$.
Each transition of $M$ that moves both heads at once is simulated by two transitions that move the two heads one after another in $M'$.  Formally, for all $q, s \in Q_M$, $x, y \in \Sigma_\bowtie$,
if $\delta_M(q, x, y) = (s, +1, +1)$, we set $\delta_{M'}(q, x, y) = (s', +1, 0)$.  Furthermore, for all $s \in Q_M$, $x, y \in \Sigma_\bowtie$, we set $\delta_{M'}(s', x, y) = (s, 0, +1)$.

If a transition of $M$ is stationary, i.e., is of the form $\delta_M(q, x, y) = (s, 0, 0)$, it is a member of either an infinite sequence representing a loop (of length  at most $\abs{Q_M}$) in which $M$ scans the symbols $x$ and $y$ without changing the head positions, or a finite sequence ending with acceptance, rejection, or the moving of some head. In the infinite-loop case, we set the corresponding transition in $M'$ to $\delta_{M'}(q, x, y) = (\qrej, +1, 0)$.
In the finite-sequence case, the value of $\delta_{M'}(q, x, y)$ will be set to $(\qacc, +1, 0)$ or $(\qrej, +1, 0)$ if the sequence is ending with acceptance or rejection, respectively, and to the value of the final transition in the sequence otherwise.

Any transition of $M$ that moves a single head is inherited by $M'$ without modification.

It may be the case that the new machine built according to these specifications moves its second head first. This problem can be handled easily by just rearranging the transition function to effectively ``swap'' the names of the two heads.
(Such a simple swap is possible, because the fact that both heads scan the left end-marker symbol at the beginning means that it is only the transition function, and not the particular input string, that determines which head moves first.)
\end{proof}


\newcommand\LIK{\ensuremath{L_{\mathit{IK}}}}

Consider the language $\LIK=\Set{a^i b^j c^k | i=j \text{ or } i=k \text{ or } j=k}$, which is in \ODFAK{3}, but not in \ODFAK{2}~\cite{IK75}. A  real-time, finite-state verifier using a single random bit can verify \LIK{} by checking certificates of the form $\sigma x^l$, where $\sigma$ is a ternary symbol that indicates which two of the three ``segments'' of the input string are claimed to be of the same length $l$. Depending on the values of $\sigma$ and the random bit, the verifier decides which  segment to attempt to match with the certificate postfix $x^l$, and accepts only if this match succeeds.

More generally, for any $k>0$, there exists a language of the form
\begin{equation*}
    L_n=\Set{y_1\#y_2\#\dotsm\#y_{2n} | y_i \in \Set{a,b}^{*} \text{ and } y_i = y_{2n+1-i}, \text{ for } 1 \leq i \leq n}
\end{equation*}
which can be recognized by
a \odfak{k+1}, but not by any \onfak{k}
\cite{YR78}. Such a language $L_n$ can be verified by a real-time, constant-space machine using $\ceil{\log (n+1)}$ random bits to split into $n+1$ paths that would compare the relevant segments of a certificate of the form $y_1\#y_2\#\dotsm\#y_{n}$ with the corresponding input segments. So we have 
$\vcscri{\rtX} \setminus \ONFAK{k} \neq \emptyset$ for all $k \geq 1 $.  


We now exhibit a language that is verifiable in real time by constant-random\-ness finite-state machines, but is unrecognizable by any deterministic multihead automaton.

\begin{theorem}\label{theorem:nonpal}
$\vcscri{\rtX} \setminus \bigcup_k \ODFAK{k} \neq \emptyset$.
\end{theorem}
\begin{proof}
We will construct a verifier $V$ for the language \nonpal{}, which was noted to be outside $\bigcup_k \ODFAK{k}$ in \cref{sec:prel}.

Every string $w$ in \nonpal{} matches the pattern $x\sigma y\sigma'z$, where $x,y,z \in \Set{0,1}^*$  and $\sigma, \sigma' \in \Set{0,1}$, such that $\abs{x}=\abs{z}$ and $\sigma \neq \sigma'$. The correct certificate $c_w$ for such an input will encode the positions of the ``unmatching'' symbols $\sigma$ and $\sigma'$ as follows:
 \begin{center} $c_w=0^{\abs{x}}10^{\abs{y}}$ \end{center}
 
$V$ tosses a single coin at the beginning of the computation to probabilistically ``branch'' to one of two ``deterministic verifiers'' $V_0$ and $V_1$, each of which checks the certificate $0^i10^j$ in a different way, as  described below.

 Note that, if  $0^i10^j$ is indeed a correct certificate for the input, claiming that the two unmatching symbols are at positions $i+1$ and $i+j+2$, then the input string must be exactly $i+1$ symbols longer than this certificate. $V_0$ checks this by moving the certificate head only once for every two moves of the input head over the input string until it passes over the 1 in the certificate. At that point, it switches to moving the certificate head at every step as well. If the certificate is of the correct length, the two heads will consume their right end-markers simultaneously, in which case $V_0$ will accept.

The task of $V_1$ is to assume that the certificate is well-formed in the sense described above, and accept if the two symbols at positions $i+1$ and $i+j+2$ really are unequal. This can be done by moving the certificate head at the same speed as the input head, recording the symbol at   position $(i+1)$ in memory, and comparing it with the input symbol scanned at the step where the certificate string has been consumed completely.

If the input is a member of \nonpal{}, both $V_0$ and $V_1$ accept with the correct certificate. 
Otherwise, the input is a palindrome, and the certificate will either be malformed (and therefore be rejected by $V_0$), or the two symbols it points out will be equal, in which case it will be rejected by $V_1$.
\end{proof}

\section{Concluding remarks}\label{sec:conc}

\newcommand\OQMARK{\textbf{\texttt{?}}}
\newcommand\LMARK{\ensuremath{\blacklozenge}}
\newcommand\LTOTMARK{\ensuremath{\bigstar}}

\newcommand\generalsets{\ensuremath{\mathrel{\bigcirc\!\!\!\bigcirc}}}

\begin{figure}[H]
    \vspace*{-1em}
    \centering
    
    \begin{tikzpicture}
        \newlength{\nodevdist}
        \newlength{\nodehdist}
        \nodevdist=14mm
        \nodehdist=33mm
        \pgfmathsetmacro\MathAxis{height("$\vcenter{}$")}
        \newcommand{\alignedeq}{%
                \sbox0{$=$}%
                \sbox0{\raisebox{-\MathAxis pt}{\usebox0}}
                \dp0=\ht0 
                \usebox0%
            }
        \begin{scope}[every node/.style={inner sep=0}, node distance=\nodevdist and \nodehdist]
            \node (1dfa2) {\ODFAK{2}};
            \node [above=of 1dfa2.center, anchor=center] (1dfa3) {\ODFAK{3}};
            \node [above=1.2\nodevdist of 1dfa3.center, anchor=center] (1dfak) {$\bigcup_k\ODFAK{k}$};
            \node [left=of 1dfak.center, anchor=center] (rtver) {\vcscri{\rtX}};
            \node [left=1\nodehdist of rtver.center, anchor=center] (1nfaklnrr) {\ONFAKLNRR{*}{1}};
            \node [right=of 1dfak.center, anchor=center] (1nfa2) {\ONFAK{2}};
            

            \node [above=1.1\nodevdist of 1dfak.center, anchor=center] (1wver1nfak) {\alignedeq};
            \node [left=20mm of 1wver1nfak.center, anchor=center] (1wver) {\vcscri{\owayX}};
            \node [right=20mm of 1wver1nfak.center, anchor=center] (1nfak) {$\bigcup_k\ONFAK{k}$};
        \end{scope}
        
        \begin{scope}[every edge quotes/.style={fill=white,sloped,allow upside down,auto=false},
						eqfn/.style={above=-.6em},
						sbfn/.style={above=-.7em},
						shorten <=1mm,
						shorten >=1mm]
            \path
                (1dfa2) edge ["$\subsetneq$"] (1nfa2)
                (1dfa3) edge ["$\supsetneq$"] (1dfa2)
                        edge ["$\generalsets$"] (1nfa2)
                (1dfak) edge ["$\supsetneq$" pos=0.63] (1dfa3)
                        edge ["$\generalsets$"] (1nfa2)
                (1nfak) edge ["$\supsetneq$"] (1nfa2)
                (rtver) edge ["$\subseteq$"] (1wver)
                        edge ["$\nsubseteq$"] (1dfak)
                        edge ["$\nsubseteq$" pos=0.27, shorten <=1.002mm, bend right=15] (1nfa2)
                        edge ["$\nsubseteq$", shorten <=1.005mm] (1dfa3)
                        edge ["$\supsetneq$", shorten <=1.003mm] (1dfa2)
                        edge ["\alignedeq"] (1nfaklnrr)
                (1wver1nfak) edge (1wver) edge (1nfak) edge ["$\supsetneq$"] (1dfak);
                
        \end{scope}
    \end{tikzpicture}
    \caption{An inclusion diagram of our results.}
    \label{fig:conclusioninclusion}
    \vspace*{-1em}
\end{figure}
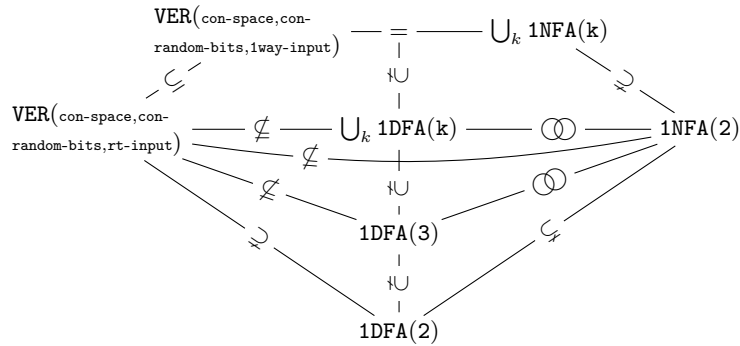

\cref{fig:conclusioninclusion} summarizes the landscape of complexity classes covered in this paper.  The \generalsets{} symbol denotes that the two related sets are neither disjoint, nor a subset of one another.


We conjecture that 
\begin{equation*}\label{eq:conj}
    \vcscri{\rtX} \subsetneq \vcscri{\owayX},
\end{equation*}
that is, restricting the input head to move in real-time yields machines which are not capable of verifying some languages that can be handled by  verifiers with one-way input. 
The reasoning behind this conjecture is based on considerations of the following languages:
\begin{align*}
    \lmatch &= \Set{x\#y_1\#y_2\#\dotsm\#y_k | \begin{array}{l}
    x, y_i \in \Set{0,1}^+ \text{ for all } i, k>0, \\
    \text{and } y_i=x \text{ for some } i
    \end{array} }\\
    L_{\frac{1}{2}} &= \Set{ww | w \in \Set{0, 1}^*}\\
    \lfrac &= \Set{xwx | x, w \in \Set{0,1}^*  \text{ and } \abs{x}=\abs{w}}
\end{align*}

These languages, which are in \vcscri{\owayX}, seem to be beyond the capabilities of real-time  verifiers. Verification of membership in these languages requires two input substrings (whose lengths are not bounded, and which cannot therefore fit in a fixed amount of memory) to be ``matched'' in a certain sense. Furthermore, the start position of the second substring cannot be determined in a one-way pass without external help. Therefore, membership certificates have to contain  information about both the position of the second substring and the content of these substrings. We suspect that it is impossible  to design certificates from which  real-time input machines can acquire these two pieces of information without getting tricked into accepting some illegal inputs.

The characterization provided by \Cref{theorem:nfakequalsver} can also be useful in the proof of the limitations of real-time  verifiers. Note that any \onfaklnrr{k}{1} $M$ that recognizes \lmatch{} seems to be forced to use an unbounded amount of nondeterminism to decide which $y_i$ substrings to try to ``match'' with the initial substring $x$. Since $M$ would have to move each head by at least one step for each nondeterministic choice it makes, sufficiently long input strings would then cause  all heads to leave $x$ behind, making a match impossible to perform.     


For further study, it would be interesting to examine the power of  real-time finite-state verifiers with less severe bounds on the amount of randomness that can be used, as well as real-time verification of debates~\cite{DSY15} between two opposing ``provers'' by similarly restricted machines. Restricting the verifiers further by imposing other conditions like reversibility~\cite{Pin92} is another possible direction.

Lastly, every non-trivial reduction naturally increases the number of states (thus the size of the program implementing it) to some extent, which is important when assessing their practicality.
A thorough analysis
in this regard is in order.




\section*{Acknowledgments}

The authors thank Martin Kutrib and Markus Holzer for  helpful discussions, and the anonymous referees for their detailed and constructive comments. This research was partially supported by Bo\u{g}azi\c{c}i
University Research Fund Grant Number 22A01P1.
Utkan Gezer's participation in this work is supported by the Turkish Directorate of Strategy and Budget under the TAM Project number 2007K12\nobreakdash-873.

\begingroup
\raggedright
\bibliographystyle{splncs04}
\bibliography{references} 
\endgroup

\end{document}